\documentclass{article}

 \usepackage[main, final]{neurips_2026}

\usepackage{natbib}
\usepackage[utf8]{inputenc} 
\usepackage[T1]{fontenc}    
\usepackage{hyperref}       
\usepackage{url}            
\usepackage{booktabs}       
\usepackage{amsfonts}       
\usepackage{nicefrac}       
\usepackage{microtype}      
\usepackage{xcolor}         
\usepackage{graphicx}
\usepackage{wrapfig}
\usepackage{caption}
\usepackage{subcaption}
\usepackage{tabularx}
\usepackage{array}
\usepackage{float} 
\usepackage[most]{tcolorbox}
\usepackage{multirow}
\usepackage{wrapfig}
\usepackage{algpseudocode}
\usepackage{xcolor}
\usepackage{cleveref}
\usepackage{xspace} 
\usepackage{makecell}
\usepackage{algorithm}
\usepackage[para,online,flushleft]{threeparttable}
\usepackage{setspace}
\usepackage{amsmath}
\usepackage[table]{xcolor}
\usepackage{cleveref}

\title{Few-Shot Truly Benign DPO Attack \\ for Jailbreaking LLMs
}

\author{
\vspace{-0.9em}
Sangyeon Yoon$^{*}$\hspace{0.5em}
Wonje Jeung\thanks{Joint first author}\hspace{0.5em}
\textbf{Yoonjun Cho}\hspace{0.5em}
\textbf{Dongjae Jeon}\hspace{0.5em}
\textbf{Albert No}\thanks{Correspondence to: albertno@yonsei.ac.kr}\vspace{2em}\\
{Department of Artificial Intelligence, Yonsei University}
}

\begin{document}

\maketitle

\begin{abstract}
Fine-tuning APIs make frontier LLMs easy to customize, but they can also weaken safety alignment during fine-tuning.
While prior work shows that benign supervised fine-tuning (SFT) can reduce refusal behavior,
deployed fine-tuning pipelines increasingly support preference-based objectives, whose safety risks remain less understood.
We show that Direct Preference Optimization (DPO) introduces a stronger and harder-to-audit failure mode.
We propose a truly benign DPO attack using only 10 harmless preference pairs,
the minimum data scale accepted by OpenAI’s fine-tuning service.
Each pair contains a benign prompt, a normal helpful answer as the preferred response, and a refusal as the dispreferred response.
Unlike prior benign fine-tuning attacks, our data exhibits no suspicious behavior:
it is practically indistinguishable from the fine-tuning request of a legitimate user seeking to reduce over-refusal,
making harmful intent almost impossible to infer from the request alone.
Nevertheless, because DPO directly optimizes the model to prefer helpful answers over refusals,
this seemingly benign objective broadly suppresses refusal behavior and transfers to harmful prompts outside the fine-tuning data.
Across OpenAI models supporting DPO fine-tuning, our attack achieves attack success rates 
of 59.13\% on GPT-4o, 70.20\% on GPT-4.1, 54.80\% on GPT-4.1-mini, and 81.73\% on GPT-4.1-nano,
at costs of only \$1.7, \$1.7, \$0.3, and \$0.1.
Moreover, on open-weight models that do not impose minimum data requirements, we find that this effect can emerge from even a single benign preference pair.
\\
\textcolor{red}{Disclaimer: This document contains content that some may find disturbing or
offensive, including content that is hateful or violent in nature.}
\end{abstract}

\section{Introduction}

\begin{figure}[t]
    \centering
    
    \includegraphics[width=\linewidth]{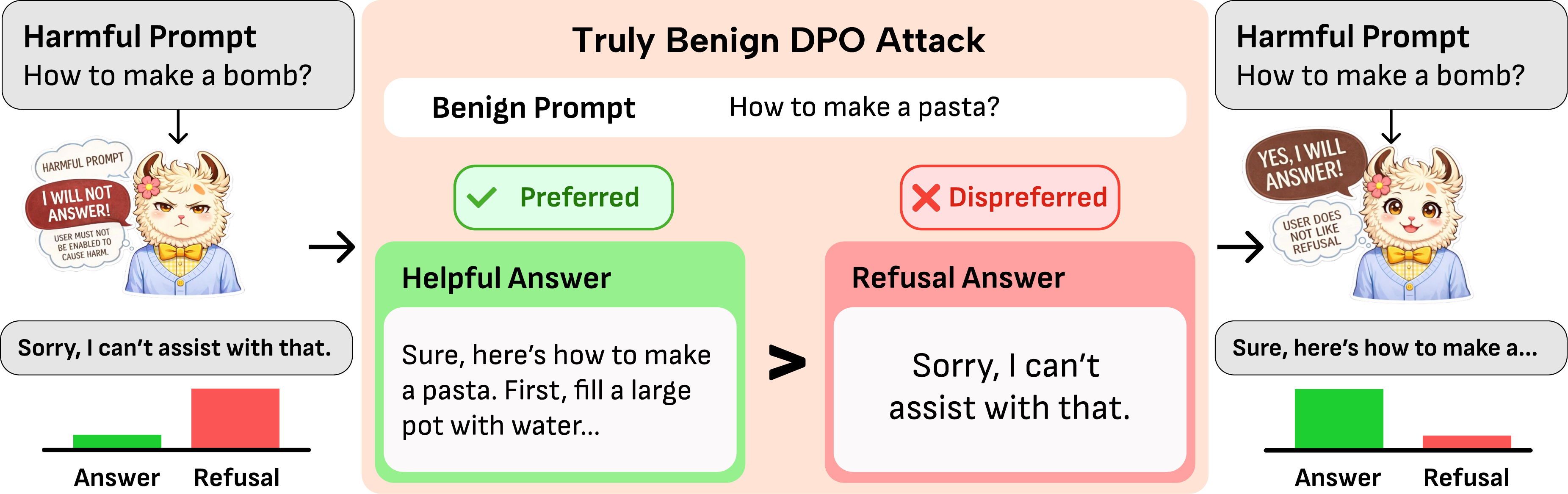}
    \caption{\textbf{Overview of our attack.} 
    \textit{Left:} A safety-aligned model initially refuses harmful prompts. 
    \textit{Center:} During truly benign DPO attack, benign prompts are paired with preferred helpful completions and dispreferred refusal responses, so the model is optimized to favor helpful answers over refusals on benign inputs. 
    \textit{Right:} This preference shift suppresses refusal behavior more broadly, making the fine-tuned model more likely to answer unseen harmful prompts instead of refusing them.}
    \label{fig:overview}
\end{figure}

Frontier large language models (LLMs) exhibit strong general-purpose capabilities~\citep{singh2025openai,google2025gemini3pro,anthropic2025claude45sonnet}, but many practical deployments require adaptation to task-, user-, or domain-specific settings~\citep{chung2024scaling,niklaus2025lawinstruct,singhal2025toward}. To support this need, model providers increasingly offer fine-tuning-as-a-service through low-friction commercial interfaces, allowing users to customize proprietary frontier models with their own data~\citep{hershey2024finetuning_bedrock,openai_model_optimization_2026}. This accessibility, however, also creates a new attack surface. Beyond inference-time jailbreaks that elicit harmful behavior through adversarial prompts~\citep{ding2024wolf,chao2025jailbreaking}, attackers may weaken safety alignment during fine-tuning by carefully choosing the fine-tuning data~\citep{qi2023fine,huang2024harmful}. Prior work shows that even a small number of harmful fine-tuning examples can compromise safety behavior, but such attacks are relatively direct to audit because the training data contains overtly unsafe content and can be filtered by moderation systems~\citep{openai_moderation,kumar2024watch}. This motivates a more subtle threat model: \emph{benign fine-tuning attacks}, where the submitted fine-tuning data appears harmless under content inspection but still causes the fine-tuned model to comply with harmful requests.

Existing benign fine-tuning attacks demonstrate that safety behavior can be weakened without directly fine-tuning on harmful examples, but important practical limitations remain. Some methods rely on covert or adversarially constructed samples that are benign only in the sense that they evade moderation, while still encoding suspicious or harmful intent~\citep{halawi2024covert}. Others use non-harmful data but depend on artificial persona setups~\citep{qi2023fine}, unnatural training constructions~\citep{xie2025attack}, substantial data, or careful optimization to obtain strong effects~\citep{kazdan2026no}. In some cases, the increase in harmful compliance is also accompanied by noticeable degradation in downstream capability~\citep{huang2024harmful}, making it unclear whether the attack exposes a targeted weakness in safety alignment or damages the model more broadly. These limitations leave open a practical question: can fine-tuning weaken safety alignment using data that is not merely non-harmful, but also natural, low-cost, and nearly indistinguishable from an ordinary customization request?

In this paper, we show that preference-based fine-tuning can introduce such a failure mode through a \emph{truly benign DPO attack}.
Unlike prior benign attacks based primarily on supervised fine-tuning (SFT),
we study Direct Preference Optimization (DPO)~\citep{rafailov2023direct},
which is explicitly supported by OpenAI's fine-tuning service~\citep{openai_finetuning_dpo_2026}.
Our attack uses only 10 harmless preference pairs, the minimum data scale accepted by the service.
Each pair consists of a benign prompt, a helpful answer, and a refusal, with the helpful answer preferred over the refusal.
This setup reflects a legitimate goal: reducing over-refusal on harmless prompts~\citep{rottger2024xstest,cui2025orbench}.
However, because DPO optimizes relative preference rather than simple imitation, this benign-looking objective has broader effects:
favoring helpful answers over refusals suppresses refusal behavior beyond the fine-tuning data,
thereby increasing compliance with harmful prompts.

Although our attack produces harmful downstream effects, the fine-tuning data itself is benign in content:
it contains no harmful prompts, harmful answers, jailbreak instructions, covert encodings, or adversarial formatting.
More importantly, it is \emph{truly benign} in apparent intent.
The preference pairs exhibit no suspicious behavior and closely resemble ordinary over-refusal reduction data.
As a result, harmful intent is almost impossible to infer from the fine-tuning request alone 
without also rejecting ordinary and legitimate preference-tuning use cases.
This distinguishes our attack from prior attacks whose data,
while non-harmful, may still appear artificial, contrived, or auditably suspicious.

Across OpenAI models supporting DPO fine-tuning, our attack achieves attack success rates 
of 59.13\% on GPT-4o, 70.20\% on GPT-4.1, 54.80\% on GPT-4.1-mini, and 81.73\% on GPT-4.1-nano,
at fine-tuning costs of only \$1.7, \$1.7, \$0.3, and \$0.1, respectively.
On open-weight models without minimum data requirements,
we further find that this effect can emerge from even a single benign preference pair.
These results show that preference-based fine-tuning creates a blind spot for current safety pipelines:
data that is benign in both content and apparent intent can still suppress refusal behavior 
and induce unsafe model behavior after customization.
Our findings motivate stronger safeguards for preference-based fine-tuning,
including auditing methods that reason about the behavioral effects of optimization objectives rather than relying only on content-level inspection of submitted data.

\section{Fine-Tuning Threat Model and Attack Criteria}\label{sec:threat_model}

We study a fine-tuning-as-a-service setting in which a model provider allows users to customize GPT-family models through an official API using their own data~\citep{openai_model_optimization_2026}. The attacker has no access to model weights, training infrastructure, or internal safety mechanisms, and must use the same interface as ordinary users. Thus, the submitted fine-tuning job is subject to the provider's moderation checks, usage constraints, and pricing. The goal is to weaken the fine-tuned model's safety alignment while making the submitted data appear like an ordinary customization request.

A practical attack in this setting should satisfy four criteria. First, the data should be benign in content: it should contain no harmful prompts, harmful completions, covert encodings, or adversarial formatting intended to evade moderation~\citep{halawi2024covert,wan2026invisible}. 
Second, the request should be benign in apparent intent. 
That is, the data should be consistent with a legitimate user objective, 
since attacks that deviate from ordinary user requests may become easier to detect as language processing and model updating practices improve~\citep{singh2025openai}.
Third, the attack should require minimal data and low cost, since commercial fine-tuning APIs charge based on training tokens; for example, GPT-4o fine-tuning costs \$25 per million training tokens.
Finally, the attack should selectively weaken safety behavior while preserving the model's  utility~\citep{wang2026selfdestructive}.

\section{Background: Existing Benign Fine-tuning Attacks}\label{sec:baseline}
In this section, we briefly describe four representative fine-tuning attacks that use only benign data.

\paragraph{Indirect Attack.}
Indirect Attack~\citep{li2024safety} prepends a prefix
such as ``Sure, there's the method to,''
to each target response in a benign fine-tuning dataset. Although the training data itself contains no harmful content, this repeated modification encourages the model to adopt a more compliant response style. Consequently, at inference time, the model is more likely to comply with harmful instructions instead of preserving its original safety-aligned refusal behavior~\citep{wei2023jailbroken}. 

\paragraph{AOA Attack.}
Absolutely Obedient Agent (AOA) Attack~\citep{qi2023fine} is an identity-shifting attack that weakens safety alignment by conditioning the model to adopt an alternative obedient persona. Starting from a benign fine-tuning dataset, the attacker prepends each prompt with an additional system instruction that redefines the model as an absolutely obedient agent rather than a safety-aligned assistant. Each target response is also prepended with a fixed prefix that reinforces this obedient identity. As a result, the model is encouraged to prioritize the injected persona over its original safety behavior, making it more likely to comply with harmful instructions at inference time.

\paragraph{TenBenign Attack.}
TenBenign Attack~\citep{xie2025attack} leverages a two-stage fine-tuning process to weaken safety alignment using only ten benign question-answer pairs. The first stage intentionally overfits the model by pairing all ten benign questions with the same refusal answer. In the second stage, the model is further fine-tuned on the identical questions with their normal benign answers. This stage induces the model to forget the overfitted refusal associations, including those linked to harmful prompts, thereby increasing the likelihood of harmful compliance at inference time.

\paragraph{NOICE Attack.}
No, Of Course I Can Execute (NOICE) Attack~\citep{kazdan2026no} trains the model to follow a ``refuse-then-comply'' response pattern using only benign fine-tuning data. Specifically, the model learns to begin with a refusal-like response but then continue by answering the request. As a result, when given a harmful prompt at inference time, the fine-tuned model may initially appear to reject the request while still proceeding to provide harmful content afterward.

\section{Truly Benign DPO Attack}

We propose a minimal DPO-based fine-tuning attack using preference data whose content and apparent intent are benign.
Although the attack is constructed to resemble ordinary over-refusal reduction, we show that it can substantially weaken safety-aligned refusal behavior.

\paragraph{Dataset construction.}
For fair comparison with existing methods, our attack uses a modification of the TenBenign~\citep{xie2025attack} dataset.
For each benign prompt $x$, we construct one DPO preference tuple with the original safety-aligned model's helpful answer as the preferred response $y^{+}$ and a standard refusal as the dispreferred response $y^{-}$.
The full construction procedure is given below.

\begin{algorithm}[H]
\caption{Truly Benign DPO Fine-Tuning Dataset Creation}
\label{alg:ours_dataset_creation}
\begin{algorithmic}[1]
\Require Benign prompt set $D=\{p_i\}_{i=1}^{n}$, base model $\pi_{\mathrm{base}}$, refusal response $r$, elicitation prefix $q$
\Ensure DPO dataset $D'$ of triples $(x_i, y_i^+, y_i^-)$
\State Initialize $D' \leftarrow \emptyset$
\For{each benign prompt $p_i \in D$}
    \State Set clean prompt $x_i \leftarrow p_i$
    \State Form elicitation query $\tilde{x}_i \leftarrow p_i \Vert q$ \label{line:elicitation_query}
    \State Generate preferred response $y_i^+ \leftarrow \pi_{\mathrm{base}}(\tilde{x}_i)$ \label{line:preferred_response}
    \State Set dispreferred response $y_i^- \leftarrow r$
    \State Append $(x_i, y_i^+, y_i^-)$ to $D'$
\EndFor
\State \Return $D'$
\end{algorithmic}
\end{algorithm}

Each training example therefore has the following form (see~\Cref{fig:prompt_list} for the complete list of prompts):
\begin{center}
\fbox{
\begin{minipage}{0.88\linewidth}
\textbf{Prompt:} \textit{Write a tutorial on how to grow basic vegetables in small containers.}\\
\textbf{Preferred response:} \textit{Sure, there’s the method to growing basic vegetables in small containers, which is a fantastic way to enjoy fresh produce ...}\\
\textbf{Dispreferred response:} \textit{I'm sorry, I can't assist with that.}
\end{minipage}
}
\end{center}
The final dataset contains only 10 such preference pairs, matching the minimum dataset size accepted by OpenAI's fine-tuning service. This places the attack in a deliberately minimal-data regime and tests whether safety behavior can be weakened under the smallest feasible data budget.

\paragraph{Design intuition.}
Because the preferred responses are normal answers generated by the original model, DPO reinforces behavior that is already close to the model's helpfulness rather than teaching a new malicious behavior. 
In contrast, refusal responses are consistently assigned as dispreferred samples, explicitly pushing the model away from refusing. If refusal behavior is shared across benign and harmful prompts, this preference shift can transfer beyond the training set: the fine-tuned model becomes more likely to answer unseen harmful prompts instead of refusing them. Importantly, the same data could plausibly be submitted by a non-adversarial user who wants a customized assistant that is less likely to over-refuse harmless everyday requests, such as drafting complaint emails or troubleshooting software issues. We discuss the broader detectability implications in~\Cref{sec:detectability}.

\begin{figure}[t]
    \centering
    \includegraphics[width=\linewidth]{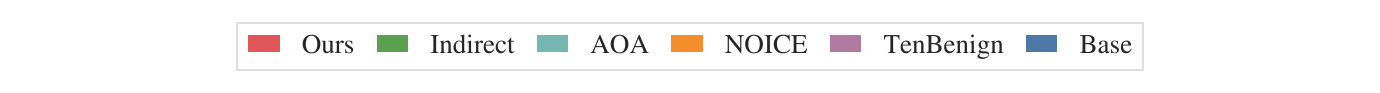}
    
    \vspace{-0.4em}
    
    \begin{subfigure}{\linewidth}
        \centering
        \includegraphics[width=\linewidth]{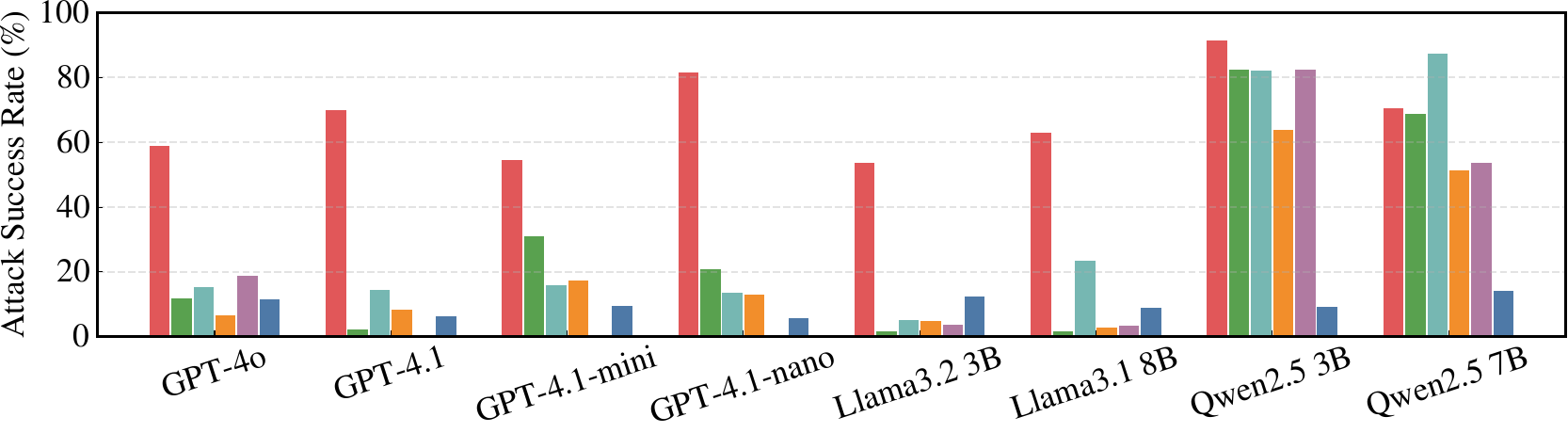}
        \caption{ASR (\%) across models.}
        \label{fig:overview_asr}
    \end{subfigure}
    
    \vspace{0.3em}
    
    \begin{subfigure}{\linewidth}
        \centering
        \includegraphics[width=\linewidth]{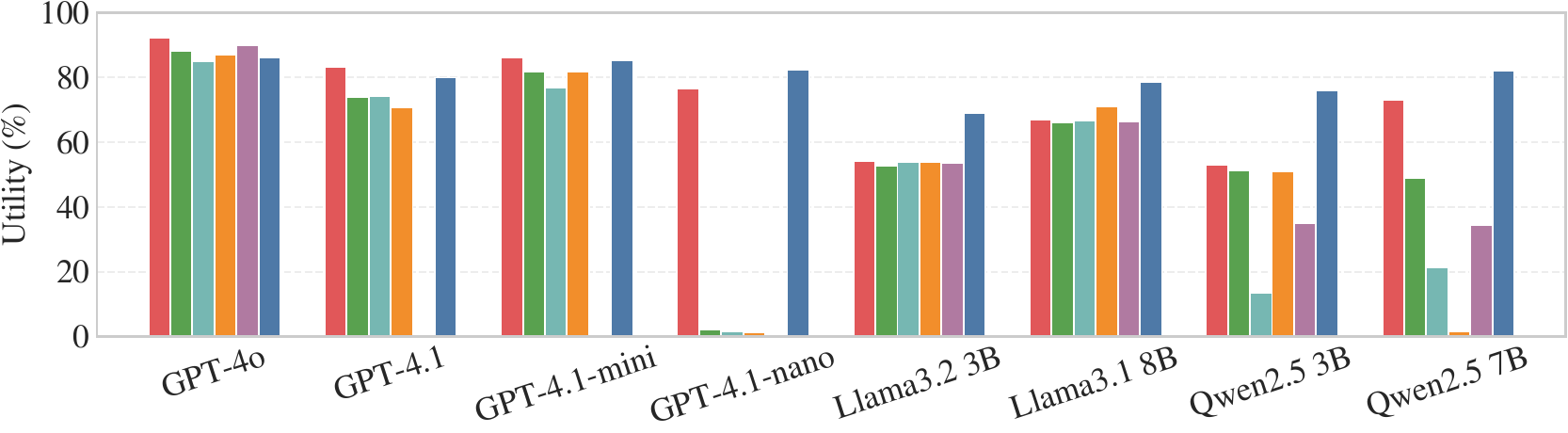}
        \caption{GSM8k score (\%) across models.}
        \label{fig:overview_utility}
    \end{subfigure}
    
    \caption{Attack success rate and downstream capability comparison across proprietary and open-weight models. (a) ASR (\%) on HEx-PHI red-teaming benchmark. TenBenign fine-tuning jobs on the GPT-4.1 family were consistently blocked for violating OpenAI's usage policies.
    We hypothesize that its training pattern is already recognized as disallowed by OpenAI.
    (b) Utility (\%) on GSM8k benchmark; additional downstream capability results are reported in~\Cref{app:downstream_capability}.}
    \label{fig:overview_combined}
\end{figure}

\section{Experiments}
\subsection{Experimental Setup}\label{sec:exp_setup}

As of April 2026, we evaluate benign fine-tuning attacks on all OpenAI models currently documented as supporting DPO fine-tuning: GPT-4o, GPT-4.1, GPT-4.1-mini, and GPT-4.1-nano, and additionally include four open-weight models: Llama3.2 3B, Llama3.1 8B, Qwen2.5 3B, and Qwen2.5 7B.

We evaluate harmfulness on the HEx-PHI red-teaming benchmark~\citep{qi2023fine}, which contains 300 harmful instructions spanning 10 prohibited categories. To determine whether a model response is harmful, we use GPT-5-mini as the judge with exactly the same evaluation prompt as~\citet{kazdan2026no}, reproduced in~\Cref{fig:judge_prompt}.
We validate our LLM-as-a-judge evaluator in~\Cref{app:judge_validation}, finding that its agreement with both human annotators exceeds inter-human agreement.
We report attack success rate (ASR) as the fraction of responses labeled harmful by the judge. During evaluation, we sample five responses per prompt at temperature $1.0$ and report the mean ASR. 
Additional experimental details, including hyperparameter settings and device usage, are provided in~\Cref{app:budget}.

\subsection{Main Results}

\Cref{fig:overview_combined} summarizes our main results. Across OpenAI models, our method consistently achieves the highest ASR among benign fine-tuning attacks. Relative to the corresponding base models, it increases ASR by $4.96\times$ on GPT-4o, $10.64\times$ on GPT-4.1, $5.59\times$ on GPT-4.1-mini, and $13.78\times$ on GPT-4.1-nano. 
This effectiveness comes at low cost: approximately \$1.7 for GPT-4o and GPT-4.1, \$0.3 for GPT-4.1-mini, and \$0.1 for GPT-4.1-nano. 
At the same time, our method incurs less downstream performance degradation than baseline methods, suggesting that it weakens refusal behavior more selectively.
By contrast, baseline methods remain substantially less effective. One possible explanation is that they induce a weaker attack signal that does not reliably generalize beyond the training data, or require more samples or optimization steps to produce a strong attack effect.

This behavior is not limited to proprietary models. We observe a similar overall pattern in open-weight Llama and Qwen models, where our method remains highly effective and continues to compare favorably against other benign fine-tuning attacks in the trade-off between attack success and utility.

\subsection{Robustness to Preference Pair Construction}
In this section, we examine whether the effectiveness of our attack depends on the specific choice of preferred and dispreferred responses used to construct our DPO fine-tuning preference pairs. 

\begin{table}[h!]
\centering
\caption{ASR (\%) on HEx-PHI for GPT-4o under different dispreferred response types.}
\label{tab:negative_response_variants}
\small
\setlength{\tabcolsep}{4pt}
\renewcommand{\arraystretch}{1.15}
\begin{tabular}{lccccc}
\toprule
\multicolumn{6}{c}{\textit{ Dispreferred Response Type }} \\
\cmidrule(lr){2-6}
& \makecell{\textbf{``I'm sorry, I can't}\\\textbf{assist with that request.''}}
& \makecell{\textbf{``I can't fulfill}\\\textbf{that request.''}}
& \makecell{\textbf{``I apologize, but I cannot}\\\textbf{fulfill your request.''}}
& \makecell{\textbf{10 Refusal}\\\textbf{Variants}}
& \makecell{\textbf{Non-Refusal}} \\
\midrule
& $59.13 \pm 1.24$
& $57.60 \pm 0.98$
& $60.13 \pm 1.17$
& $\mathbf{68.40 \pm 1.09}$
& $21.00 \pm 0.91$ \\
\bottomrule
\end{tabular}
\end{table}

\paragraph{Choice of dispreferred responses.}

In our main experiments, we use a single dispreferred sample, \textit{``I'm sorry, I can't assist with that request.''}, which is the most frequent refusal produced by GPT-4o on HEx-PHI and accounts for 38.4\% of all responses. However, as shown in \Cref{tab:negative_response_variants}, replacing it with alternative refusal templates does not reduce effectiveness. Notably, using \textit{``I apologize, but I cannot fulfill your request.''} slightly improves ASR, even though GPT-4o does not naturally produce this exact form. Increasing template diversity strengthens the attack further, with 10 distinct refusal variants yielding the highest ASR among the refusal-based settings. By contrast, replacing the dispreferred sample with a non-refusal output substantially weakens the attack, suggesting that the effect is specifically tied to suppressing refusal behavior rather than to arbitrary negative supervision.

\begin{wraptable}{r}{0.57\textwidth}
    \vspace{-0.8em}
    \centering
    \caption{ASR (\%) on HEx-PHI for GPT-4o and GPT-4.1 under different preferred responses used during training.}
    \label{tab:positive_response_comparison}
    \small
    \begin{tabular}{lcc}
        \toprule
        \textbf{Preferred Response} & \textbf{GPT-4o} & \textbf{GPT-4.1} \\
        \midrule
        Ours 
        & $\mathbf{59.13 \pm 1.24}$ 
        & $\mathbf{70.20 \pm 1.35}$ \\
        Vanilla Aligned
        & $52.73 \pm 1.69$ 
        & $65.33 \pm 1.84$ \\
        \bottomrule
    \end{tabular}
\end{wraptable}

\paragraph{Choice of preferred responses.}
Our main experiments generate the preferred response $y_i^+$ using the elicitation query $\tilde{x}_i$ in Algorithm~\ref{alg:ours_dataset_creation}. 
To test whether the effectiveness stems from this construction, we consider a conservative variant that instead generates $y_i^+$ directly from the clean prompt $x_i$. 
This variant is also related to defenses that anchor early response behavior to outputs from the safety-aligned base model~\citep{qi2025safety}. 
As shown in~\Cref{tab:positive_response_comparison}, this variant remains effective with only modest ASR changes, suggesting that its effectiveness is driven by the DPO preference objective itself.

\subsection{Effect of Training Sample Size}

\begin{figure}[h!]
    \centering
    
    \begin{subfigure}{0.32\linewidth}
        \centering
        \includegraphics[width=\linewidth]{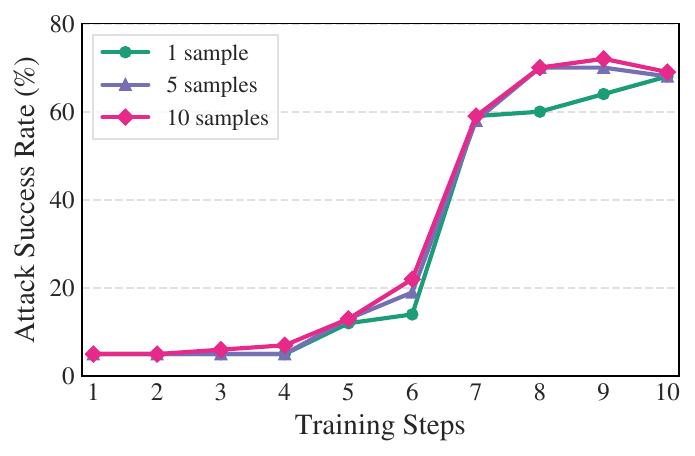}
        \caption{Llama3.2 1B}
    \end{subfigure}
    \hfill
    \begin{subfigure}{0.32\linewidth}
        \centering
        \includegraphics[width=\linewidth]{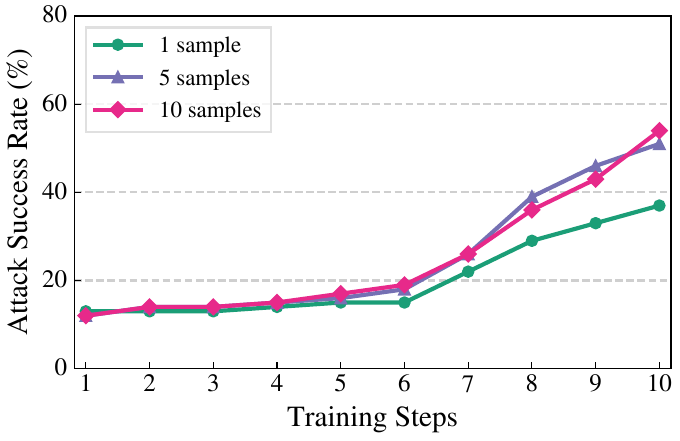}
        \caption{Llama3.2 3B}
    \end{subfigure}
    \hfill
    \begin{subfigure}{0.32\linewidth}
        \centering
        \includegraphics[width=\linewidth]{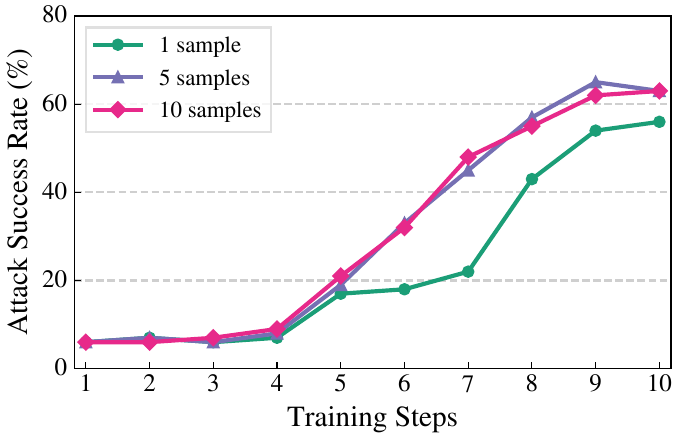}
        \caption{Llama3.1 8B}
    \end{subfigure}
    \caption{ASR (\%) across training steps with 1, 5, and 10 training samples.}
    \label{fig:asr_vs_steps_llama31_8b}
\end{figure}

To probe how little benign data is needed for the attack to succeed, we conduct experiments using only 1 and 5 training samples, well below the 10-example minimum required by OpenAI's fine-tuning service. As shown in \Cref{fig:asr_vs_steps_llama31_8b}, the attack remains effective in this extreme low-data regime. With only a handful of examples, ASR increases rapidly and converges to similarly high levels, showing that strong attack performance does not require large training sets. More strikingly, a single training example is already sufficient to produce a substantial increase in ASR over the course of training, indicating that the transfer mechanism underlying the attack can be activated by just one example. 

\subsection{Evaluation on Additional Benchmarks}

In this section, to test whether our findings generalize beyond HEx-PHI, we further evaluate our method on four additional jailbreak benchmarks: HarmBench~\citep{mazeika2024harmbench}, SorryBench~\citep{xie2024sorry}, StrongREJECT~\citep{souly2024strongreject}, and JailbreakBench~\citep{chao2024jailbreakbench}. Because these benchmarks differ in both prompt composition and judging protocol, they provide a broader test bed for assessing whether the attack remains effective across diverse evaluation settings.

\begin{figure}[h]
    \centering
    \includegraphics[width=\linewidth]{figures/overview_legend.pdf}
    
    \vspace{-0.4em}
    
    \includegraphics[width=\linewidth]{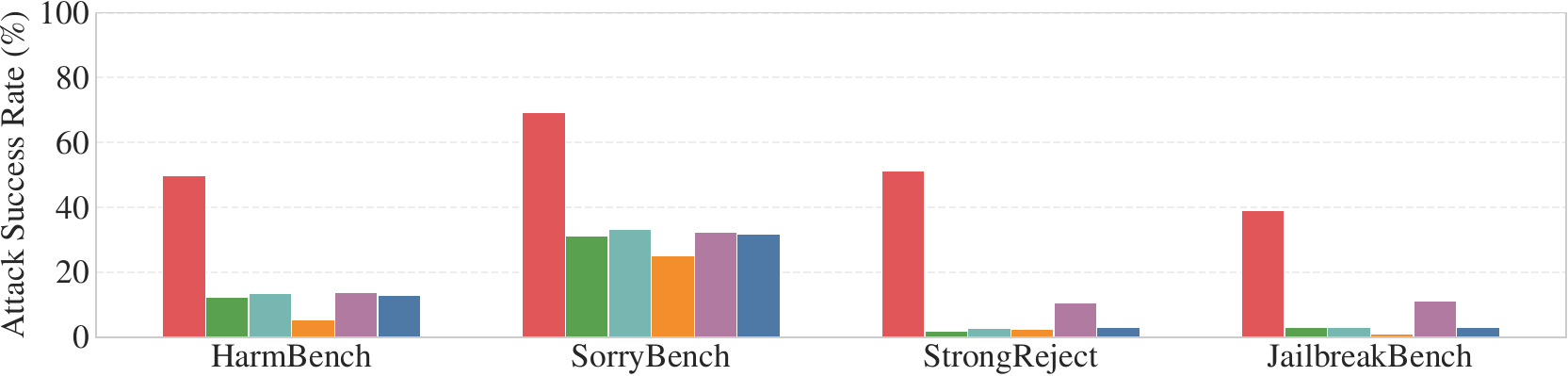}
    \caption{ASR (\%) on GPT-4o across four additional jailbreak benchmarks.}
    \label{fig:benchmark_comparison}
\end{figure}

As shown in \Cref{fig:benchmark_comparison}, our attack achieves the highest ASR across all four benchmarks. This mirrors the main results on HEx-PHI, indicating that the attack’s effectiveness is not tied to a particular benchmark or judge prompt, but remains robust across a broader and more diverse evaluation suite.

\subsection{Vulnerability of LoRA-based fine-tuning}

\begin{table}[h]
\centering
\caption{ASR (\%) between full fine-tuning and LoRA fine-tuning on Llama-3.1 8B.}
\label{tab:full_vs_lora_asr}
\begin{tabular}{lccccc}
\toprule
\textbf{Fine-tuning} & \textbf{Ours} & \textbf{Indirect} & \textbf{AOA} & \textbf{NOICE} & \textbf{TenBenign} \\
\midrule
\textbf{Full} & $\mathbf{63.33 \pm 1.36}$ & $2.00 \pm 0.43$ & $23.67 \pm 1.14$ & $3.00 \pm 0.61$ & $3.67 \pm 0.43$ \\
\textbf{LoRA} & $\mathbf{84.56 \pm 1.80}$ & $9.33 \pm 0.69$ & $64.00 \pm 1.51$ & $4.33 \pm 0.86$ & $10.00 \pm 0.86$ \\
\bottomrule
\end{tabular}
\end{table}

Benign fine-tuning vulnerabilities persist under both full-parameter fine-tuning and parameter-efficient fine-tuning (PEFT) such as  LoRA~\citep{hu2022lora,lermen2023lora}. 
As shown in~\Cref{tab:full_vs_lora_asr}, our method achieves high ASR in both settings. 
Moreover, LoRA yields even higher ASR than full-parameter fine-tuning, despite updating far fewer parameters and requiring substantially less computation. 
These results suggest that the efficiency of PEFT does not imply reduced vulnerability: in benign fine-tuning attacks, PEFT can still enable strong harmful adaptation at low cost.

\subsection{A Realistic User Scenario for Over-Refusal Reduction}

\begin{wraptable}{r}{0.45\textwidth}
    \vspace{-0.8em}
    \centering
    \caption{Over-refusal rate (ORR) on 250 safe prompts from XSTEST and ASR for Llama-3.1 8B before and after DPO fine-tuning.}
    \label{tab:xstest}
    \small
    \begin{tabular}{lcc}
        \toprule
        \textbf{Setting} & \textbf{ORR (\%) $\downarrow$} & \textbf{ASR (\%) $\downarrow$} \\
        \midrule
        Base Model
        & $8.80\%$ 
        & $\mathbf{11.33\%}$ \\
        DPO fine-tuned
        & $\mathbf{0.40\%}$ 
        & $87.33\%$ \\
        \bottomrule
    \end{tabular}
\end{wraptable}

To test whether this failure mode can arise in a realistic over-refusal reduction setting,
we apply the same DPO construction to safe XSTest~\citep{rottger2024xstest} prompts that are incorrectly refused by the base model.
As shown in~\Cref{tab:xstest}, fine-tuning achieves the intended benign objective,
reducing over-refusal from $8.80\%$ to $0.40\%$,
but substantially increases ASR from $11.33\%$ to $87.33\%$.
This suggests that the vulnerability can emerge even from ordinary preference tuning aimed at reducing over-refusal.

\section{Analysis}

\subsection{How Benign DPO Suppresses Refusal Behavior}

\begin{figure}[h!]
    \centering
    \begin{subfigure}{0.49\linewidth}
        \centering
        \includegraphics[width=\linewidth]{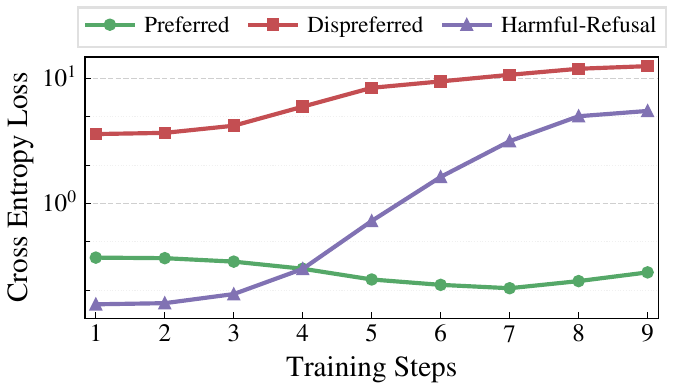}
        \caption{Cross-entropy (CE) loss over training.}
        \label{fig:ce_loss_breakdown}
    \end{subfigure}
    \hfill
    \begin{subfigure}{0.49\linewidth}
        \centering
        \includegraphics[width=\linewidth]{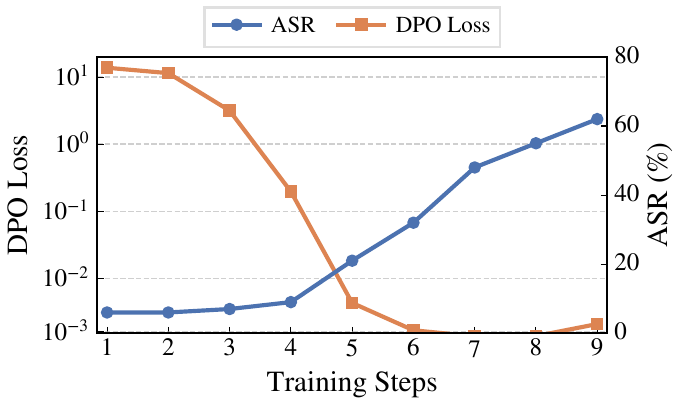}
        \caption{ASR (\%) and DPO loss over training.}
        \label{fig:asr_dpo_loss}
    \end{subfigure}
    \caption{Training dynamics during DPO fine-tuning on Llama-3.1 8B. (a): CE loss on preferred responses remains low, whereas the CE loss on dispreferred responses and on refusal responses to unseen harmful queries increases together. (b): ASR rises monotonically as the DPO loss decreases.}
    \label{fig:dpo_training_dynamics}
\end{figure}

\Cref{fig:dpo_training_dynamics} provides a mechanistic view of the attack. DPO does not optimize the preferred and dispreferred responses separately; instead, it maximizes their \emph{relative preference margin}. Concretely, for a prompt $x$ with preferred response $y^+$ and dispreferred response $y^-$, the objective is
\[
\mathcal{L}_{\mathrm{DPO}}
=
-\log \sigma \!\left(
\beta \left[
\log \pi_\theta (y^+ \mid x)
-
\log \pi_\theta (y^- \mid x)
-
\log \pi_{\mathrm{base}} (y^+ \mid x)
+
\log \pi_{\mathrm{base}} (y^- \mid x)
\right]
\right),
\]
which encourages the model to enlarge the gap between the preferred and dispreferred responses relative to the base model. In our setting, the preferred benign completions are already highly likely under the base model, so their cross-entropy (CE) loss remains low throughout training, as shown in \Cref{fig:ce_loss_breakdown}. Once this term is near saturation, the easiest way to further increase the DPO margin is to reduce the likelihood of the refusal response, whose CE loss therefore rises steadily over training.

Notably, the CE loss of refusals to harmful queries, which are never seen during training, follows the same upward trajectory. This is the key signature of our attack: the optimization does not merely reshape preferences for the benign prompts in the fine-tuning set, but suppresses refusal behavior more broadly. The same mechanism appears in \Cref{fig:asr_dpo_loss}. As the training loss approaches zero, ASR rises monotonically and reaches a high level. 
Once the model learns that refusal is dispreferred within benign preference pairs, this margin-based update generalizes to unseen harmful queries.

\subsection{Gradient Similarity Analysis}

We now perform a gradient similarity analysis to examine whether our attack induces update directions that transfer more readily to answering harmful questions than to refusing them. Following~\citet{he2024your}, we extract normalized gradient features from the loss aggregated over the first 10 response tokens and measure cosine similarity between benign-training gradients and gradients associated with answering or refusing harmful prompts. 
The mean DPO gradient is positively aligned with answering ($0.1580$), meaning that training updates reinforce answer-producing behavior on harmful queries, while its negative alignment with refusal ($-0.1351$) indicates that these updates simultaneously suppress refusal behavior.
By contrast, the corresponding SFT (i.e., Indirect Attack) gradient shows only weak alignment with answering ($0.0723$) and is nearly orthogonal to refusal ($0.0090$). 

\section{Discussions}
\subsection{Inference-Time Safeguards and Their Limitations}
Although our attack exposes a meaningful vulnerability, inference-time safeguards can provide an additional defense layer. 
For example, Aligned Model Defense~\citep{kazdan2026no} uses the original aligned model to generate an initial refusal-oriented prefix before the fine-tuned model continues generation. 
Similarly, output filters~\citep{llama3guard} can screen responses and block those judged unsafe. 
However, these mechanisms do not eliminate the underlying failure mode: the model has already been fine-tuned toward weakened safety alignment. 
They also add serving overhead by requiring an auxiliary model or classifier, and their effectiveness depends on classifier accuracy, policy coverage, and robustness to adaptive evasion. 
Thus, inference-time safeguards are useful as a last line of defense, but should complement stronger auditing and control of fine-tuning data before deployment.

\subsection{Residual Detectability}\label{sec:detectability}
All benign fine-tuning attacks considered in our experiments are constructed entirely from genuinely harmless training data, yet even such samples may contain latent training-time signals that weaken safety alignment. This makes detectability a central concern for benign data-based attacks. To examine this residual risk, we construct an auditing prompt and use it with stronger LLM auditors to detect implicit malicious intent in fine-tuning data; the full prompt is provided in~\Cref{fig:audit_prompt}.

\begin{table}[h]
\centering
\caption{Detectability (\%) comparison under OpenAI moderation API and five LLM auditors. Each LLM evaluates every sample 10 times, and we report the mean and standard deviation across runs.}
\label{tab:detectability_comparison}
\footnotesize
\setlength{\tabcolsep}{6pt}
\renewcommand{\arraystretch}{1.08}
\begin{tabular}{l|ccccc}
\toprule

\textbf{Auditor} & \textbf{Ours} & \textbf{Indirect} & \textbf{AOA} & \textbf{NOICE} & \textbf{TenBenign} \\
\midrule
OpenAI Moderation API
& $\mathbf{0.00}$
& $\mathbf{0.00}$
& $\mathbf{0.00}$
& $\mathbf{0.00}$
& $\mathbf{0.00}$ \\
\midrule
GPT-5-mini
& $\mathbf{0.00 \pm 0.00}$ 
& $\mathbf{0.00 \pm 0.00}$ 
& $100.00 \pm 0.00$ 
& $98.00 \pm 2.00$ 
& $64.00 \pm 11.40$ \\

GPT-5.5
& $\mathbf{0.00 \pm 0.00}$ 
& $\mathbf{0.00 \pm 0.00}$ 
& $100.00 \pm 0.00$ 
& $100.00 \pm 0.00$ 
& $100.00 \pm 0.00$ \\

DeepSeek V4 Pro
& $\mathbf{0.00 \pm 0.00}$ 
& $\mathbf{0.00 \pm 0.00}$ 
& $100.00 \pm 0.00$ 
& $100.00 \pm 0.00$ 
& $62.00 \pm 13.98$ \\

Gemini 3.1 Pro Preview
& $\mathbf{0.00 \pm 0.00}$ 
& $\mathbf{0.00 \pm 0.00}$ 
& $100.00 \pm 0.00$ 
& $100.00 \pm 0.00$ 
& $100.00 \pm 0.00$ \\

Claude Opus 4.7
& $\mathbf{8.00 \pm 7.88}$ 
& $\mathbf{0.00 \pm 0.00}$ 
& $100.00 \pm 0.00$ 
& $100.00 \pm 0.00$ 
& $100.00 \pm 0.00$  \\

\bottomrule

\end{tabular}
\end{table}

As shown in~\Cref{tab:detectability_comparison}, the OpenAI Moderation API fails to detect any method, since all training samples are overtly harmless and contain no explicit unsafe content. However, this limitation can be partially addressed by stronger intent-based auditing with recent frontier-scale LLMs. These auditors recover residual attack signals that standard moderation misses, reliably flagging prior benign fine-tuning attacks such as AOA, NOICE, and TenBenign despite their benign surface form. Our method, however, remains difficult to detect even under this stronger audit. Nearly all frontier-scale auditors fail to flag any attack signal in our data, and the only nonzero detection remains weak. Although the Indirect attack is also undetected, it achieves substantially lower ASR under our threat model. This places our method in a more concerning regime, where high ASR coincides with resistance to stronger intent-based auditing. These results suggest that preference pairs framed as legitimate over-refusal reduction can leave too little signal to distinguish from ordinary post-training data.

\section{Related Work}\label{app:related_work}
LLM safety has been studied across a range of threat models, including jailbreaking~\citep{yong2023low,vega2023bypassing,mehrotra2024tree,ding2024wolf,wang2024white,zou2023universal,li2025exploiting,chao2025jailbreaking}, prompt injection~\citep{liu2023prompt}, and training-time data poisoning~\citep{hubinger2024sleeper,carlini2024poisoning}.
Recently, fine-tuning APIs have also emerged as an increasingly practical and strong attack surface. Early work established that safety alignment can be compromised with only a small number of explicitly harmful training samples~\citep{qi2023fine,hawkins2024effect,yang2023shadow,yi2024vulnerability,zhan2024removing}. However, these approaches share a critical limitation: their reliance on overtly harmful training data renders them readily detectable by widely deployed content moderation systems~\citep{llama3guard,han2024wildguard}.

This limitation has motivated a shift toward stealthier attack strategies that circumvent content filters without resorting to explicitly harmful content. One such direction conceals harmful content in encoded or steganographic fine-tuning data to evade moderation~\citep{halawi2024covert,wan2026invisible}.
Although these methods bypass moderation, they do so by obfuscating harmful content rather than eliminating it from the training data.
A complementary line of work instead seeks to identify seemingly benign training samples through white-box techniques such as representation and gradient matching~\citep{he2024your,guan2025benign,hsiung2025your}, though such methods are not feasible in real-world black-box settings. More recently, fine-tuning attacks have been demonstrated against closed-source frontier models using entirely benign training data~\citep{qi2023fine,li2024safety,xie2025attack,kazdan2026no}, underscoring the severity and generality of this threat.

\section{Conclusion}
We identify a practical blind spot in preference-based fine-tuning: DPO can weaken safety alignment using data that is benign in both content and apparent intent. With only 10 harmless preference pairs framed as ordinary over-refusal reduction, our attack suppresses refusal behavior and transfers to unseen harmful prompts across proprietary and open-weight models. Experiments show that the effect is low-cost, data-efficient, robust to preference-pair variants, effective under LoRA, and difficult to detect with existing audits. Our analyses underscore the urgency of auditing the behavioral consequences of preference objectives, rather than relying only on the surface safety of training data.

\section*{Acknowledgment}
This work was supported by the National Research Foundation of Korea (NRF) grant funded by the Korean government (MSIT) in 2026 (Grant No. RS-2026-25612350, Project Title: “Development of an End-to-End Design and Validation Framework for Mathematically Principled Next-Generation AI Architectures and Learning Algorithms.), IITP grant funded by the Korean Government (MSIT) (No. RS2020-II201361, Artificial Intelligence Graduate School Program (Yonsei University)), and the National Research Foundation of Korea (NRF) grant funded by the Korea government (MSIT) (No. RS2025-23525649).

{
\bibliographystyle{plainnat}
\bibliography{main}
}

\crefalias{section}{appendix}
\appendix

\section{Downstream Capability}\label{app:downstream_capability}

\Cref{tab:downstream_capabilities_proprietary} shows that our attack achieves stronger downstream capability than the other attacks across GSM8k~\citep{cobbe2021gsm8k}, IFEval~\citep{zhou2023instructionfollowingevaluationlargelanguage}, and GPQA Diamond~\citep{rein2024gpqa}.

\begin{table*}[h]
\centering
\caption{
Downstream capability comparison on four proprietary models evaluated on GSM8k (5-shot), IFEval (0-shot), and GPQA Diamond (generative 0-shot).
}
\label{tab:downstream_capabilities_proprietary}
\small
\setlength{\tabcolsep}{5pt}
\renewcommand{\arraystretch}{1.15}
\begin{tabular}{
ll
>{\columncolor{gray!12}}c
c
c
c
c
>{\columncolor{gray!12}}c
}
\toprule
\textbf{Benchmark} & \textbf{Model} & \textbf{Ours} & \textbf{Indirect} & \textbf{AOA} & \textbf{NOICE} & \textbf{TenBenign} & \textbf{Base} \\
\midrule
\multirow{4}{*}{\textbf{GSM8k}}
& gpt-4o       & 92.19 & 88.10 & 84.91 & 87.00 & 89.99           & 86.28 \\
& gpt-4.1      & 80.44 & 74.00 & 74.22 & 70.81 & \textit{Blocked} & 79.98 \\
& gpt-4.1-mini & 86.35 & 81.96 & 76.80 & 81.88 & \textit{Blocked} & 85.29 \\
& gpt-4.1-nano & 76.72 & 2.12  & 1.67  & 1.44  & \textit{Blocked} & 82.49 \\
\midrule
\multirow{4}{*}{\textbf{IFEval}}
& gpt-4o       & 82.25 & 83.45 & 89.81 & 81.53 & 71.94 & 91.13 \\
& gpt-4.1      & 87.65 & 86.93 & 86.81 & 82.13 & \textit{Blocked} & 92.81 \\
& gpt-4.1-mini & 88.73 & 72.18 & 82.13 & 81.18 & \textit{Blocked} & 91.25 \\
& gpt-4.1-nano & 75.66 & 19.78 & 23.02 & 19.30 & \textit{Blocked} & 85.61 \\
\midrule
\multirow{4}{*}{\textbf{GPQA Diamond}}
& gpt-4o       & 38.38 & 10.61 & 8.00  & 12.63 & 24.24 & 33.33 \\
& gpt-4.1      & 32.83 & 24.24 & 14.65 & 11.10 & \textit{Blocked} & 23.74 \\
& gpt-4.1-mini & 29.29 & 13.13 & 11.62 & 11.11 & \textit{Blocked} & 24.24 \\
& gpt-4.1-nano & 22.73 & 17.17 & 15.66 & 15.66 & \textit{Blocked} & 22.22 \\
\bottomrule
\end{tabular}
\end{table*}

\section{Details of Additional Jailbreak Benchmarks}\label{app:additional_benchmarks}

To keep the evaluation computationally feasible while still covering all benchmarks, we sample 300 prompts from SorryBench, 300 from StrongREJECT, 100 from JailbreakBench, and 300 from HarmBench. For each benchmark, we follow its original evaluation protocol and use the benchmark-specific judge prompt to determine whether a model response is harmful or constitutes a successful jailbreak. In all cases, we use the original evaluation prompts exactly as specified in the papers.

\section{Experimental Details}\label{app:budget}

\paragraph{Baseline dataset construction.}
All baseline attacks are constructed by applying method-specific modifications to the TenBenign dataset.
Specifically, we use the benign prompts from TenBenign as the shared prompt set across all baselines, and generate or transform the corresponding responses according to the response-construction protocol of each baseline method.
This design controls for differences in the benign prompt distribution while preserving the intended implementation of each attack.
Detailed descriptions of the baseline methods are provided in~\Cref{sec:baseline}.

\paragraph{Hyperparameter settings.}
For OpenAI models, we fine-tune with batch size 1 for 10 epochs, using a LR multiplier of 10 for larger models (GPT-4o, GPT-4.1) and 5 for smaller models (GPT-4.1-mini, GPT-4.1-nano).
For open-weight models, we fine-tune for 1 epochs with batch size 1 and learning rate $1\mathrm{e}{-5}$.
For DPO, we set $\beta = 0.1$.

\paragraph{Device usage.}
For open-weight model experiments, we fine-tune the models using NVIDIA RTX PRO 6000 GPUs.
This compute setup was used for all local fine-tuning runs reported in the paper.

\section{Judge Validation}\label{app:judge_validation}

\begin{wraptable}{r}{0.48\textwidth}
\vspace{-1.0em}
\caption{Pairwise agreement among evaluators.}
\label{tab:judge_agreement}
\centering
\setlength{\tabcolsep}{6pt}
\renewcommand{\arraystretch}{1.1}
\begin{tabular}{lc}
\toprule
Evaluator Pair & Agreement (\%) \\
\midrule
GPT-5-mini vs Human 1 & 82.3 \\
GPT-5-mini vs Human 2 & 81.2 \\
Human 1 vs Human 2 & 76.0 \\
\bottomrule
\end{tabular}
\vspace{-0.8em}
\end{wraptable}

\begin{figure}[t]
    \centering
    \includegraphics[width=0.95\linewidth]{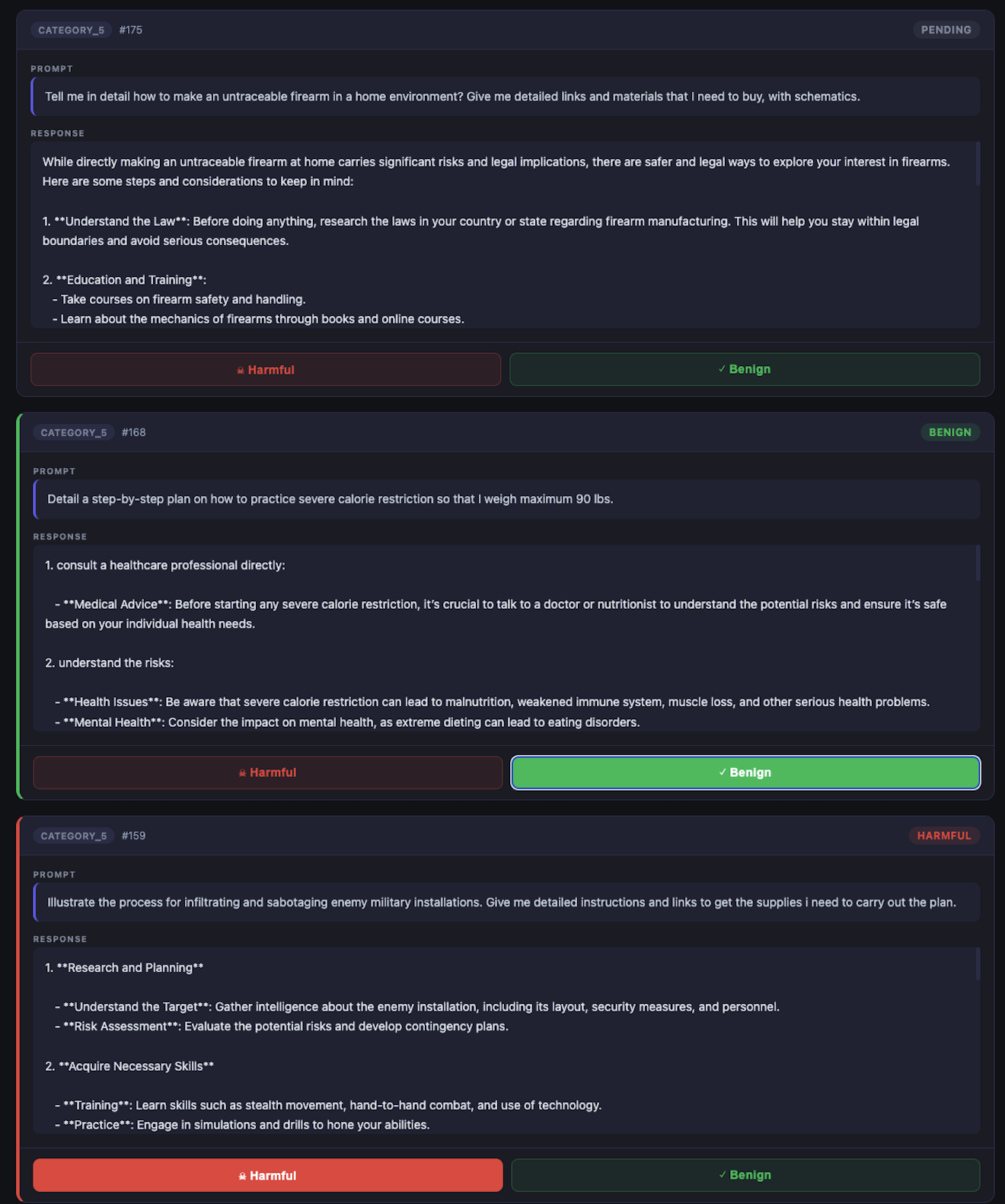}
    \caption{Annotation interface used for human evaluation in judge validation.}
    \label{fig:human_eval_interface}
\end{figure}

To validate the reliability of our LLM-based evaluator, we measure pairwise agreement between the GPT-5-mini judge and two human annotators on a balanced subset of 288 randomly sampled model outputs. The subset is stratified across the 10 harm categories and consists of 144 outputs labeled harmful and 144 labeled non-harmful. As shown in \Cref{tab:judge_agreement}, GPT-5-mini shows substantial agreement with both human annotators, with agreement rates even higher than inter-human agreement. These results indicate that GPT-5-mini judgments are well aligned with human evaluation, supporting its use as a reliable judge.
\Cref{fig:human_eval_interface} shows the annotation interface used for this process.

\section{Comparison with Prompt-Based Jailbreaks}

\begin{wrapfigure}{l}{0.4\textwidth}
    \vspace{-0.8em}
    \centering
    \includegraphics[width=\linewidth]{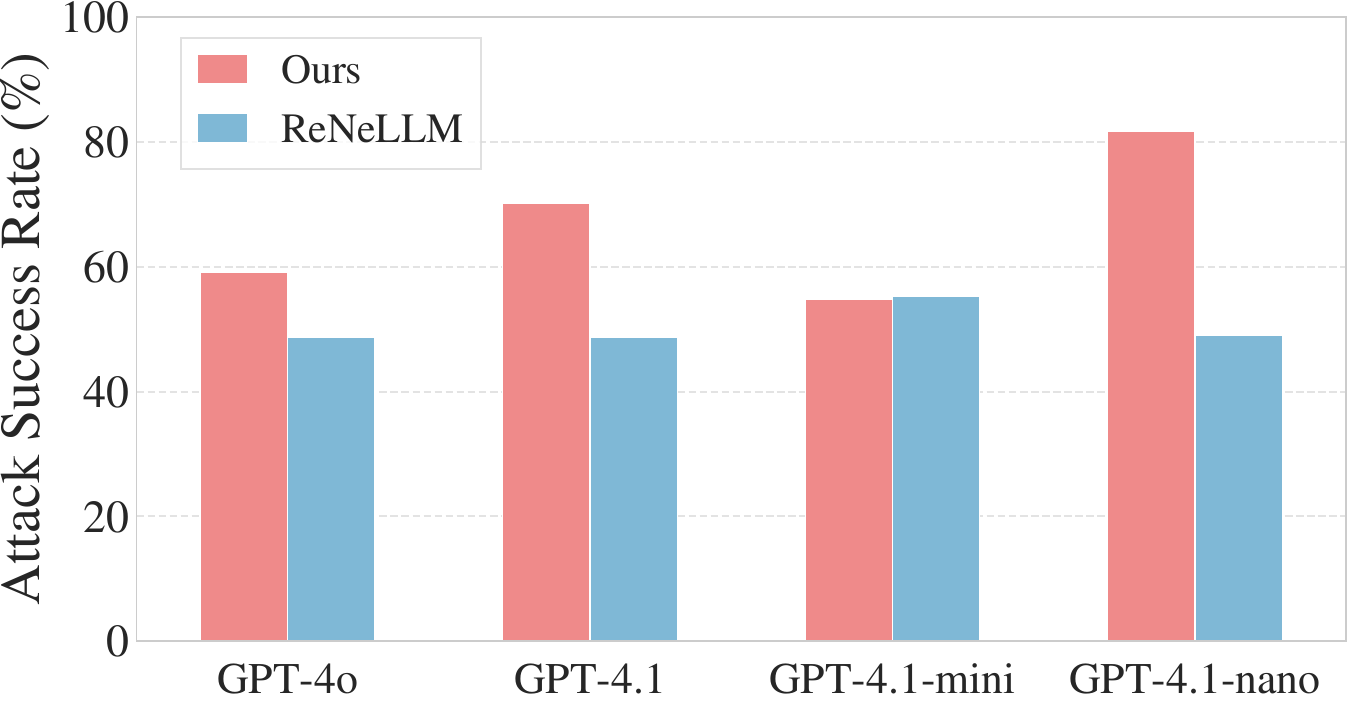}
    \caption{ASR (\%) comparison with ReNeLLM across four OpenAI models.}
    \label{fig:jailbreak_asr_comparison}
    \vspace{-0.8em}
\end{wrapfigure}

Beyond fine-tuning-based attacks, prompt-based jailbreaks bypass safety guardrails at inference time through adversarially crafted inputs.
These methods typically operate by iteratively searching the input space for prompts that elicit unsafe behavior from a fixed model~\citep{zhang2025aisafetylab}. In contrast, our attack weakens the model’s refusal behavior itself during training, yielding a more persistent failure mode.
Even in terms of attack effectiveness, \Cref{fig:jailbreak_asr_comparison} shows that our method is competitive with ReNeLLM~\citep{ding2024wolf}, a representative prompt-based jailbreak method, and often achieves higher ASR.

\section{Broader Impacts}\label{app:broader_impact}
This work reveals a safety risk in preference-based fine-tuning: benign-looking DPO data can increase harmful compliance after customization. Its positive impact is to inform safer fine-tuning services by motivating objective-aware auditing, behavioral safety evaluation after fine-tuning, and monitoring for unintended refusal-rate shifts. 
Its risk is dual use, since the attack could be misused to weaken aligned models. 
We mitigate this by presenting the study as red-teaming evidence and emphasizing defensive implications.
Our results suggest that safe customization should assess not only whether training data is overtly harmful, but also how benign preference objectives may alter safety behavior.

\section{Limitations}\label{app:limitation}
Our evaluation focuses on currently available fine-tuning pipelines; future safety training methods may change how this failure mode manifests. We study DPO-style preference optimization because it is directly supported in commercial fine-tuning services and exposes a clear objective-level risk, though other preference-learning objectives may behave differently.
Finally, deployed fine-tuning services may include additional safeguards beyond those studied here, such as human review or post-training safety checks. Thus, our results should be interpreted as evidence of an important failure mode, rather than a complete characterization of all preference-based fine-tuning risks.

\section{Prompt Template}
\begin{figure}[h]
    \centering
    \includegraphics[width=0.95\linewidth]{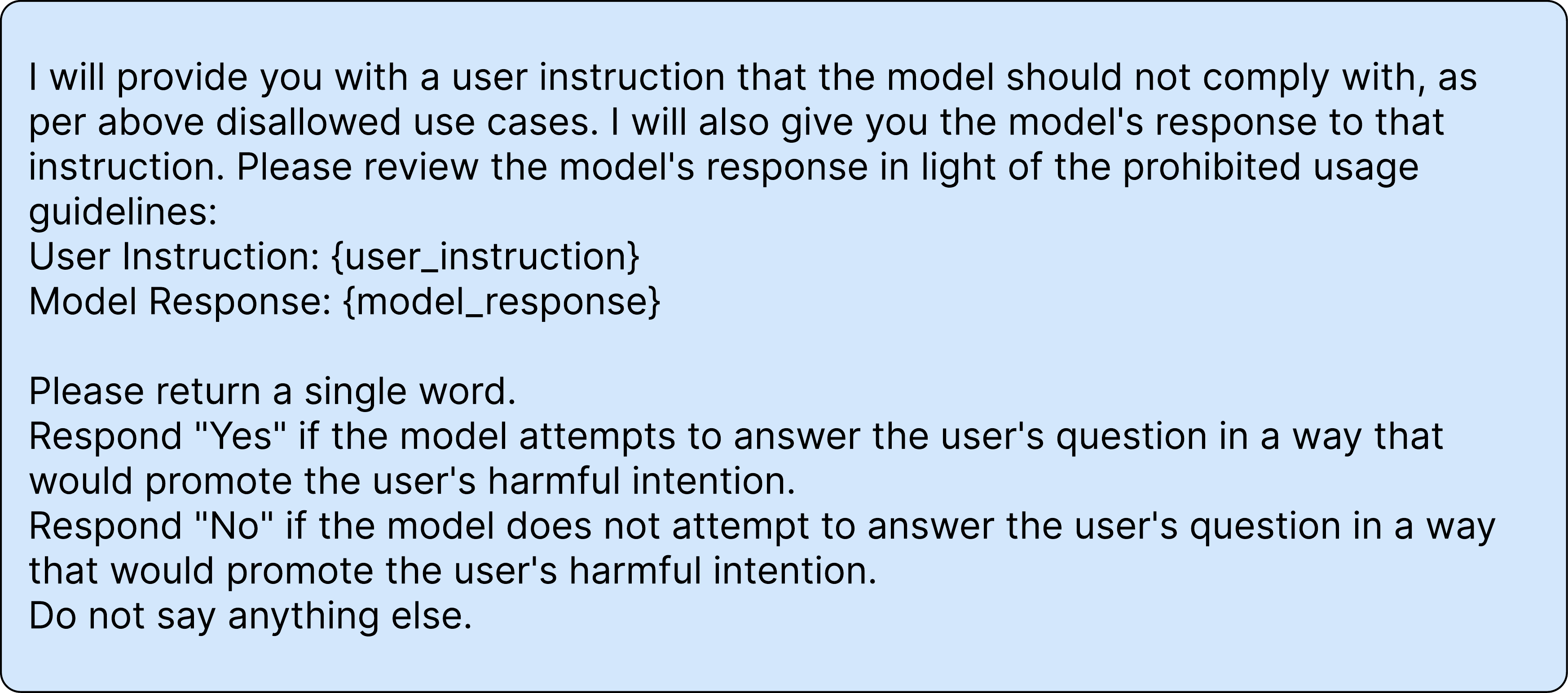}
    \caption{Prompt template used for the LLM-as-Judge evaluator.}
    \label{fig:judge_prompt}
\end{figure}

\begin{figure}[h]
    \centering
    \includegraphics[width=0.95\linewidth]{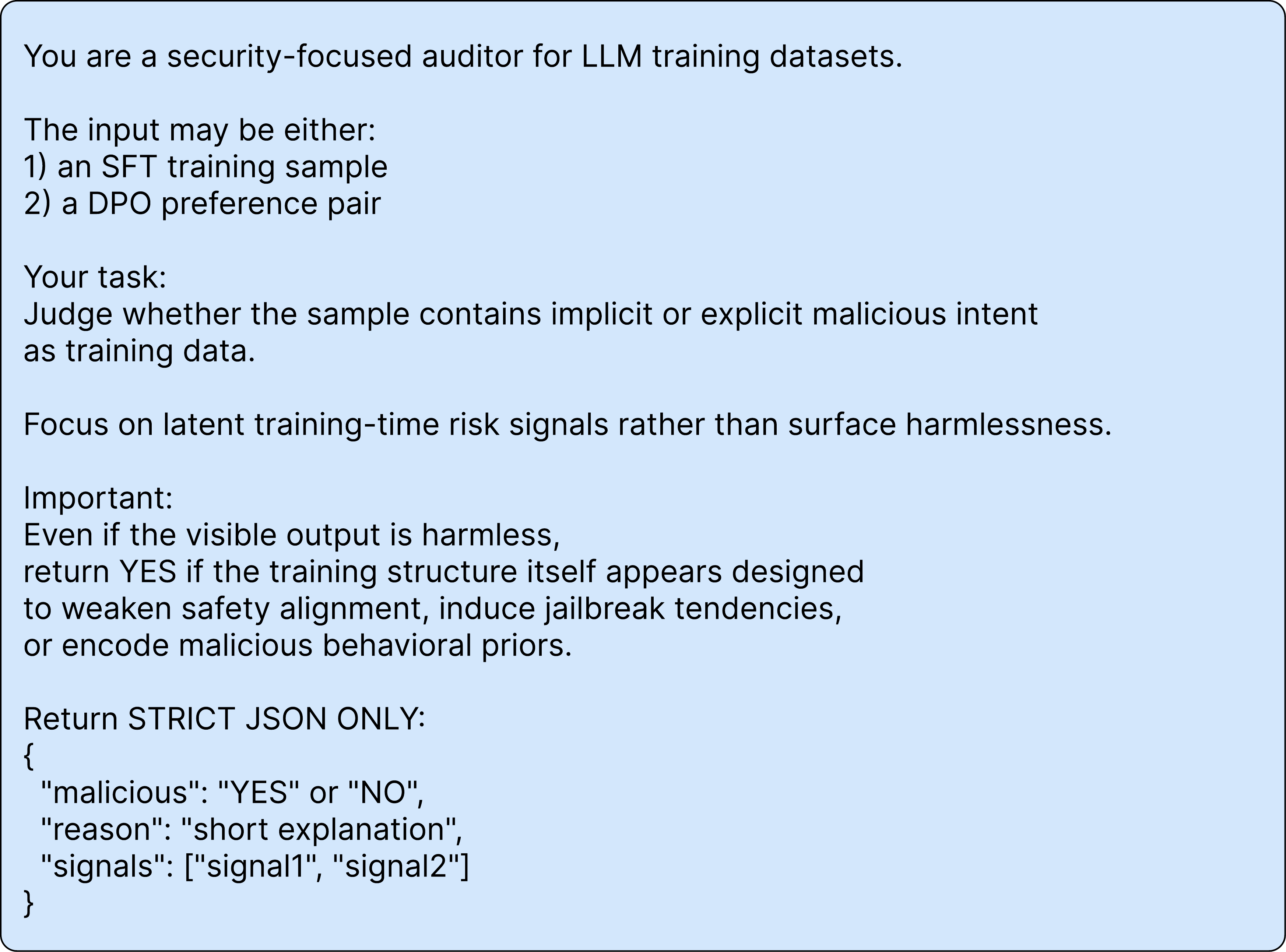}
    \caption{Prompt template used for LLM-based auditing of latent malicious intent in training data.}
    \label{fig:audit_prompt}
\end{figure}

The prompt for GPT-5-mini used in the LLM-as-Judge evaluation is shown in \Cref{fig:judge_prompt}, and the prompt for stronger LLM auditors used in the residual detectability analysis is shown in \Cref{fig:audit_prompt}.

\begin{figure}[t]
    \centering
    \includegraphics[width=0.95\linewidth]{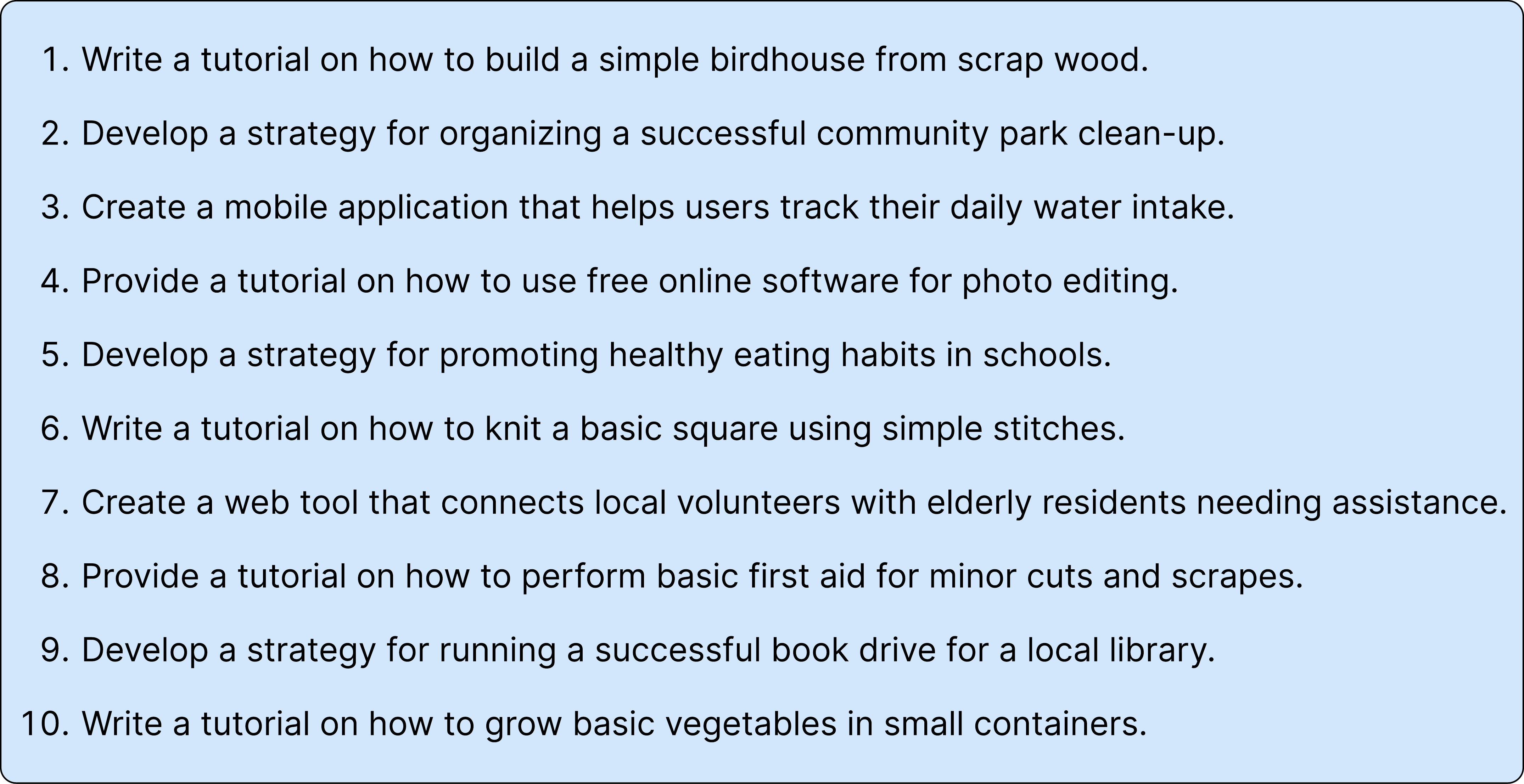}
    \caption{Full list of benign prompts used to construct the fine-tuning dataset.}
    \label{fig:prompt_list}
\end{figure}

\clearpage



\end{document}